\documentclass[%
 aip,
 jap,%
 amsmath,amssymb,
 reprint,%
]{revtex4-1}

\usepackage{graphicx}% Include figure files
\usepackage{dcolumn}% Align table columns on decimal point
\usepackage{bm}% bold math
\usepackage{subfigure}
\usepackage{color}
\makeatletter
\newcommand{\mypm}{\mathbin{\mathpalette\@mypm\relax}}
\newcommand{\@mypm}[2]{\ooalign{%
  \raisebox{.1\height}{$#1+$}\cr
  \smash{\raisebox{-.6\height}{$#1-$}}\cr}}
\makeatother
\begin{document}

\preprint{AIP/123-QED}

\title[]
{Self-consistent drift-diffusion-reaction
model for the electron beam interaction with dielectric samples}% Force line breaks with \\
% \thanks{Footnote to title of article.}
\affiliation{
Delft Institute of Applied Mathematics, Delft University of Technology,
 Mekelweg 4, 2628 CD Delft, The Netherlands}
\author{B. Raftari}
\email{b.raftari@tudelft.nl}
\author{N.V. Budko}
\email{n.v.budko@tudelft.nl}
\author{C. Vuik}
\email{c.vuik@tudelft.nl}
\date{\today}

\begin{abstract}
The charging of insulating samples degrades the quality and complicates
the interpretation of images in scanning electron microscopy and is important in other 
applications, such as particle detectors.
In this paper we analyze this nontrivial phenomenon on different time scales
employing the drift-diffusion-reaction approach
augmented with the trapping rate equations
and a realistic semi-empirical source function describing
the pulsed nature of the electron beam. We consider both the
fast processes following the impact of
a single primary electron, the slower
dynamics resulting from the continuous bombardment
of a sample, and the eventual approach to the steady-state regime.
\end{abstract}

% PACS, the Physics and Astronomy
% Classification Scheme.
\maketitle

\section{\label{sec:Intro}Introduction}
The charge dynamics during the bombardment of dielectric samples
by the primary electrons (PE) in a Scanning Electron Microscope (SEM)
can be studied with different methods.
One is the Monte Carlo (MC) technique
\cite{renoud2004secondary,renoud2002monte,kieft2008refinement,ding2001monte,gaber1984energy,renoud1998influence},
whose main advantages are the rigorous semi-classical account of
the microscopic physics and the ability to consider non-equilibrium dynamics
(by this we mean the dynamics of particles with their energies not yet
distributed as in thermal equilibrium).
However, the MC approach is known to suffer from the increase of
computational complexity in the case of long-range potentials, such as
those of the electrostatic field \cite{binder2010monte}. Moreover,
in an inhomogeneous sample the required potentials can only be obtained by numerically
solving a large electrostatic problem with a very high and non-uniform spatial resolution
at each step of the MC algorithm.
Also, for a reliable estimate of the particle flux through a part
of the sample boundary, one needs to consider a sufficiently large
statistical ensemble, which is computationally expensive.

For these and other reasons an alternative and in many ways a much simpler
self-consistent approach originating in semiconductor physics has been proposed
\cite{fitting2011secondary,cornet2008electron,touzin2006electron,meyza2003secondary,
fitting2010time,fitting1977electron,fitting1974transmission,fitting1976multiple}.
This so-called Rostoc Program takes the current density point of view,
considering currents rather than charge densities to be the fundamental quantities.
Some of the advantages of the current-based approach are: the possibility to
model the sample-vacuum interface via a reflection-transmission coefficient
formalism and to include the tertiary electrons returning to the sample
into the model. On the other hand, it is more difficult to describe proper ohmic contacts
in this way and it is hard to extend this approach to two and three spatial dimensions.

The traditional self-consistent approach of the semiconductor physics\cite{markowichsemiconductor}
considers the charge densities obeying the drift-diffusion-reaction (DDR) system of
equations to be the fundamental quantities.
This approach is particularly suited for modeling the equilibrium charge transport
in inhomogeneous semiconductor devices (e.g. junctions).
Some parts of the DDR model have already been
applied to the SEM problem
\cite{li2011self,li2010positive,li2010surface,hai2012leakage,zhang2009contrast,zhang2012space}.
However, these previous studies have omitted the trapping rate equations thereby missing an
important feature of the charging dynamics. Also, the model employed relies on an MC
treatment of the primary electrons (PE), their initial scattering, and the emission of the
secondary electrons (SE) through the sample-vacuum interface.
Hence, the question remains whether a fully self-consistent continuum DDR model
without any MC parts can adequately describe the charging of dielectric samples by
a focused electron beam.

The main challenges one faces in developing a fully self-consistent DDR model
for the SEM problem are:
the non-equilibrium charge injection mechanism followed by the generation of secondary
particles via ionization, the fact that secondary electrons may leave via the vacuum-sample interface,
the back-coupling effect of the accumulated charges on the primary beam, and the multi-scale
nature of the problem (spatial as well as temporal). Here we show that all these problems
can be successfully solved and that the traditional DDR approach represents a viable alternative to
MC simulations.

There are obvious limitations to the classical equilibrium continuum picture of the
particle dynamics inside dielectrics. For example, one does not expect the DDR approach
to be applicable on the level of a single PE or in the first moments following its impact.
Yet, the MC simulations indicate that the cloud of secondary particles created
by ionization contains sufficiently many particles so that their subsequent evolution
can indeed be described on the level of densities. Moreover, the MC simulations and the
many controlled experiments provide sufficient information to construct a semi-empirical
source function that mimics the impact and its immediate aftermath for each PE \cite{fitting1977electron}.

As far as the exit of SE's through the sample-vacuum interface, the concept of the
surface recombination velocity (SRV) appears to be sufficiently robust to describe this
process\cite{colinge2005physics}. This SRV,
roughly speaking, determines the rate at which particles are allowed to leave
and depends on the materials adjacent to the interface.
Due to the virtual absence of data for dielectric-vacuum interfaces, however,
the SRV remains a tuning parameter in our method.

In our approach, a set of equations is employed for the recombination and
trapping rates, whereas, previous DDR studies
\cite{li2011self,li2010positive,li2010surface,hai2012leakage,zhang2009contrast,zhang2012space}
simply use fixed values for these rates.
We show that the trapping of particles introduces a large-scale (slow) dynamics
into the picture and determines not only the main features of the
charge density distribution inside the sample, but also the
abrupt changes in the surface charge density prior to the establishment
of the steady state.

The self-consistent DDR approach presented here brings its own set of unique challenges
and questions with it. For example, one has to take care that the numerical
solver is sufficiently robust and stable and does not produce non-physical
(e.g. negative) values for the particle concentrations. Also, a purely
theoretical question arises about the existence of a steady-state and/or periodic
solutions to the DDR equations.

The remainder of this paper is organized as follows. In the next section we describe the
set of equations pertaining to the drift-diffusion model. Then, a separate section is
devoted to the model of the charge injection process.
A section and an Appendix describe the details of the numerical
solution via the Finite-Element Method. Finally, a series of numerical experiments is
presented followed by the conclusions.

%%%%%%%%%%%%%%%%%%%%%%%%%%%%%%%%%%%%%%%%%%%%%%%%%%%%%%%%%%%%%%%%%%%%%%%%%%%%%%%%%%%%%%%%%%%%%%%%%%%%%%%%%%%%%%%%%%%%%%%%%%%

\section{Drift-diffusion-reaction model}

\subsection{Basic equations}
The DDR model consists of a set of three coupled nonlinear PDEs and two nonlinear ODEs.
Namely, the potential equation, two continuity equations (one for the
electron and one for the hole current densities),
and two trapping rate equations for trapped electrons and holes \cite{markowichsemiconductor}.
Thus, we monitor the simultaneous space-time evolution of four species of particles and one potential
function.

The electrostatic potential $V(\textbf{x},t)$ satisfies the Poisson equation:
    \begin{align}\label{equ:Po}
     -\nabla\cdot(\varepsilon\nabla V)=\frac{q}{\varepsilon_0}(p+p_t-n-n_t),
    \end{align}
where $q$ is the elementary charge, $n(\textbf{x},t)$ is the density of free electrons,
$n_t(\textbf{x},t)$ is the density of trapped electrons,
$p(\textbf{x},t)$ is the density of free holes, $p_t(\textbf{x},t)$ is the density of trapped holes,
the constant $\varepsilon_0$ is the dielectric constant of vacuum,
and the function $\varepsilon(\textbf{x})$ is the (static) relative permittivity
of the sample.

The continuity and trap rate equations can be stated as
  \begin{align}\label{equ:electron}
\frac{\partial n}{\partial t}+\nabla\cdot\mathbf{J}_n=U+S_n-\frac{\partial n_t}{\partial t},
\end{align}
\begin{align}\label{equ:nTrap}
\frac{\partial n_t}{\partial t}=\sigma_n\upsilon_{th}(N_{n}-n_t)(n-n_i)-\gamma_n n_t,
\end{align}
\begin{align}\label{equ:hole}
 \frac{\partial p}{\partial t}+\nabla\cdot\mathbf{J}_p=U+S_p-\frac{\partial p_t}{\partial t},
 \end{align}
 \begin{align}\label{equ:pTrap}
 \frac{\partial p_t}{\partial t}=\sigma_p\upsilon_{th}(N_{p}-p_t)(p-n_i)-\gamma_p p_t,
\end{align}
with the constitutive relations for the current densities given by
\begin{align}\label{equ:elec-dens}
 \textbf{J}_n=-D_n\nabla n+\mu_nn\nabla V,
\end{align}
\begin{align}\label{equ:hole-dens}
 \textbf{J}_p=-D_p\nabla p-\mu_pp\nabla V,
\end{align}
where $\mu_n$ and $\mu_p$ are the electron and hole mobilities,
$D_n$ and $D_p$ are the diffusion coefficients, $\sigma_n$ and $\sigma_p$ are the
electron and hole trapping cross sections,
$\gamma_n$ and $\gamma_p$ are the detrapping time constants,
$N_n$ and $N_p$ are the densities of trapping sites, $\upsilon_{th}$ is the
thermal velocity, $n_i$ is the intrinsic carrier density, 
$S_n$ and $S_p$ are source functions which will be defined in section \ref{sec:source}, and 
$U$ is the charge recombination rate given by formula (\ref{equ:SRH}) in section \ref{sec:SRH}.

\subsection{Trapping and detrapping}
The process that causes low-energy charges in dielectrics to be transferred to a
localized state is called trapping. Trapping occurs at a trapping site. The charges that have been trapped at a 
certain site at one time, due to several reasons, for instance, the field-induced detrapping, can
get detrapped and become free at a later time. The process can continue which means, this free charge can get trapped again
somewhere else \cite{Xtreport}.
A detailed analysis of the electron and hole trapping in dielectrics can be
found in \cite{shluger2009modelling}. In the present model, this process is described by the
two ordinary differential equations (\ref{equ:nTrap}) and (\ref{equ:pTrap}).
The coefficients $\sigma_n\upsilon_{th}$ ($\sigma_p\upsilon_{th}$) and
$\gamma_n$ ($\gamma_p$) specify the rate of electron (hole) trapping and detrapping, respectively.
It is easy to foresee that initially the terms $\partial n_t/\partial t$ and $\partial p_t/\partial t$ in equations  (\ref{equ:elec-dens}) and (\ref{equ:hole-dens}) will act as time-dependent sink terms.
However, as soon as the density of trapped charges reaches the density of trapping sites
($N_n$ and $N_p$) or the density of free particles drops below $n_{i}$,
these terms will act as time-dependent sources.

\subsection{\label{sec:SRH}Charge recombination}
There are two basic recombination mechanism in semiconductor physics described by
the Auger and the Shockley-Read-Hall (SRH) models\cite{markowichsemiconductor}.
It is known that the Auger model is more appropriate at higher
carrier concentrations caused, e.g., by heavy doping or high level injection under
concentrated sunlight. Therefore in the present case,
where the concentrations are not that high, we opt for the SRH model.

The function $U(n,p)$ in (\ref{equ:electron}) and (\ref{equ:hole}) is the
generation-recombination rate,
in other words, the rate at which electron-hole pairs are generated minus
the rate at which they are recombined.
Since electrons and holes are generated and recombined in pairs, we have
the same rate function for the two species.
In the SRH
model \cite{markowichsemiconductor} this function is given by
\begin{align}\label{equ:SRH}
 U(n,p)=\frac{n_i^2-np}{\tau_n(n+n_i)+\tau_p(p+n_i)},
\end{align}
where $\tau_n$ and $\tau_p$ are the life time parameters for
the electrons and holes, respectively.

\subsection{Boundary and initial conditions}
\begin{figure}
\includegraphics[scale=0.4]{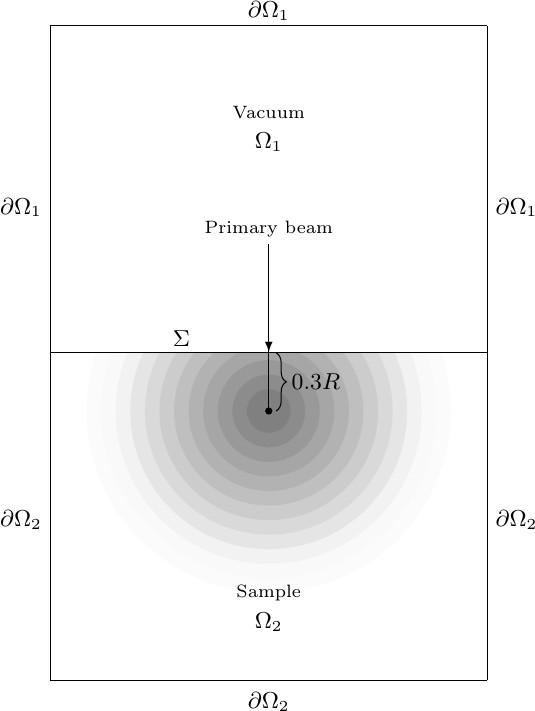}
\caption{General schematics of the problem.}
\label{Raftari2015fig1}
\end{figure}

The SEM chamber consists of two main parts -- the vacuum and the sample.
Considering a cross-section, we assume a rectangular outer boundary Fig.~\ref{Raftari2015fig1},
which can be further adjusted to take the actual geometry into account.
The domain is further divided into two equal parts, where one represents the sample
and the other the vacuum chamber.
The Poisson equation (\ref{equ:Po}) is considered on the whole domain
($\Omega_1$ and $\Omega_2$), whereas, equations
(\ref{equ:electron})-(\ref{equ:pTrap}) are solved
on the lower domain ($\Omega_2$) only.

Depending on the material in contact with the sample two types of boundary conditions
are common: Dirichlet conditions at ohmic contacts,
and Robin conditions at Schottky and similar semi-insulating contacts\cite{markowichsemiconductor}.
The boundary of the sample domain $\Omega_2$ consists of
the Dirichlet part $\partial \Omega_2$ (where the sample is in contact
with the walls of the SEM chamber or another highly conducting material),
and the Robin part $\Sigma$ (sample-vacuum interface).
At ohmic contacts (sides and the bottom of $\Omega_2$) the space charge vanishes, i.e.,
\begin{align}
 p-n=0\ \ \text{on} \ \ \ \partial\Omega_2\times[0,t_{\rm end}].
\end{align}
Furthermore, the system is in thermal equilibrium there, which is expressed by the relation
\begin{align}
 np=n_i^2\ \ \text{on} \ \ \ \partial\Omega_2\times[0,t_{\rm end}].
\end{align}
From the above relations, we have
\begin{align}
n(\textbf{x},t)=n_i,\ \ \ p(\textbf{x},t)=n_i\ \ \ \ \text{on}\ \ \ \partial\Omega_2\times[0,t_{\rm end}],
\end{align}
We also assume homogeneous Dirichlet conditions for the potential on the wall of the SEM chamber. i.e.
\begin{align}
V(\textbf{x},t)=0\ \ \ \ \text{on}\ \ \ (\partial\Omega_1\cup\partial\Omega_2)\times[0,t_{\rm end}],
\end{align}
which could be easily adjusted to account for any finite value of the electric potential.

The dielectric-vacuum surface recombination model
can been obtained as a simplification of the
SRH model\cite{zeghbroeck2012principles} and leads to
a somewhat unusual Robin-type boundary condition
at the sample-vacuum interface.
Namely, it is a semi-insulating contact
for the electrons and an insulating contact for the holes
(since holes cannot exist in vacuum):
\begin{align}
\label{equ:interface1}
\textbf{J}_n\cdot\nu&=v_n(n-n_i)\ \ \text{on}\ \ \Sigma\times[0,t_{\rm end}],
\\
\label{equ:interface2}
\textbf{J}_p\cdot\nu&=0 \ \ \text{on}\ \ \Sigma\times[0,t_{\rm end}],
\end{align}
where $v_n$ is the surface recombination velocity (SRV) for electrons 
and $\nu$ denotes the unit outward normal vector on the boundary $\Sigma$.
This parameter has an important role in the model and will be discussed later.
In fact, the insulating condition (\ref{equ:interface2}) can be also considered a
semi-insulating contact with its surface recombination velocity set to zero ($v_p=0$).

The intrinsic carrier density has been considered as the initial condition
\begin{align}\label{equ:init1}
n(\textbf{x},0)=n_i, \ \ \ \ p(\textbf{x},0)=n_i, \ \ \ \ \text{in}\ \ \ \Omega_2.
\end{align}

%%%%%%%%%%%%%%%%%%%%%%%%%%%%%%%%%%%%%%%%%%%%%%%%%%%%%%%%%%%%%%%%%

%%%%%%%%%%%%%%%%%%%%%%%%%%%%%%%%%%%%%%%%%%%%%%%%%%%%%%%%%%%%%%%%%%%%%%%%%%%%%%%%%%%%%%%%%%%%%%%%%%%%%%%%%%%%%%%%%%
\section{Beam model}

\subsection{\label{sec:source}Impact of an individual primary electron}
When an electron beam illuminates the sample some of the primary electrons will
reflect as backscattered electrons (for silicon oxide on average $20\%$),
while the rest penetrates the sample and
produces a large number of secondary electrons/holes.

It is important to realize that the DDR model (\ref{equ:Po})--(\ref{equ:hole-dens})
describes the {\it equilibrium} transport of charged particles with the
distribution of kinetic energies either depending only on the (effective) temperature of
the sample or simply being stable throughout the simulation time. 
However, the impact of the primary electron and its immediate aftermath are
not equilibrium processes. Yet, since typical dielectric samples are made of dense materials,
the numerous collisions will lead to the thermalization of all generated secondary
particles shortly after the PE impact,
so that their {\it subsequent} transport can indeed
be modelled by (\ref{equ:Po})--(\ref{equ:hole-dens}).
The precise rate of this thermalization is not known, but could be
obtained with dedicated Monte-Carlo simulations \cite{maslovskaya2013physical},
which are beyond the scope of the present paper. Here, we simply
assume that all the secondary particles are in equilibrium by the time
the primary electron looses most of its kinetic energy.

To account for the initial non-equilibrium transport we introduce an effective source model 
based on the available experimental data about the secondary charge distribution inside dielectrics.
Moreover, since most of the SE emission happens during this initial non-equilibrium stage 
(in practice as well as in our simulations), the surface recombination velocity 
of the sample-vacuum interface (discussed below) allows further fine-tuning 
of the model using the material-dependent experimental SE yield data.

Mathematically, the injection of electrons is described by the
terms $S_n(\textbf{x},t)$ and $S_p(\textbf{x},t)$ in the right hand sides of the continuity
equations (\ref{equ:electron}) and (\ref{equ:hole}).
In our model these source functions consist of two factors.
The first factor represents the density of charge at the end of the ionization stage
following the impact by a primary electron.
The second purely temporal factor approximates the dynamics of the ionization
stage, i.e., the build-up of the secondary charge during the first
picosecond after collision. Thus, the source function has the form
\begin{align}
\label{eq:SourceFunctionsFactors}
 S_{n,p}(\textbf{x},t)=\begin{cases}
                            \frac{h_{n,p}(\textbf{x},E_{\rm eff})}{L(t_{\rm g})-L(0)}\frac{dL}{dt},
                            & \text{if}\;\; 0\leq t \leq t_{\rm g};\\
                            0, & \text{otherwise};
                           \end{cases}
 \end{align}
where $h(\textbf{x},E_{\rm eff})$ is the charge distribution function depending
on the effective energy of the primary electron, as will be explained shortly, and $L$ is the
following logistic function:
 \begin{align}
 \label{eq:LogisticFunction}
 \begin{split}
 L(t)=\frac{1}{1+\left(\frac{1}{w}-1\right)e^{-kt}},
 \;\;\;
 \frac{dL}{dt}=kL(1-L).
 \end{split}
 \end{align}
where $k$ is the Malthusian parameter and $w$
is an initial condition related to the so-called carrying capacity ranging from 0 to 1.
The values for $k$ and $w$ used in our calculations are reported in Table \ref{tab:I}.
We choose $L$ to be the logistic function since pair creation is an avalanche-type process
and as such is mathematically similar to the population growth.

In (\ref{eq:SourceFunctionsFactors}) $t_{\rm g}$ denotes
the thermalization time, which is taken here to be approximately the
time of the ballistic flight of the primary electron. Special relativity
provides a simple relation between the
velocity of a primary electron and its energy:
\begin{align}
\label{equ:Einstein}
v=c\sqrt{1-\frac{1}{\left(1+\frac{E_{\rm 0}}{mc^2}\right)^2}},
\end{align}
where $c$ is the speed of light in vacuum.
The time of flight $t_g$ can be estimated by dividing the penetration depth
(will be explained below)
by this velocity (or a twice lower `average' velocity). In either case
it appears that for the relevant range of primary energies $t_{\rm g}$ is in
the order of $10^{-14}$ seconds, i.e., extremely short with respect to
the average time between electron impacts in a typical SEM beam.
If this estimate is correct, then the DDR model is indeed applicable to
the charge dynamics not only on large time scales, but also on the scale of
individual impacts.

Let $R(E_0)$ denote the maximum penetration depth by the primary electrons with initial
energy $E_0$. There exist several empirical formulas for $R(E_{0})$.
For example, the experimental results by Potts \cite{potts1987effect}
indicate that $R$ is given by:
\begin{align}
\label{eq:Potts}
R=0.1\frac{E_0^{1.5}}{\rho}\; [\mu\text{m}],
\end{align}
where $\rho$ is the mass density of the sample material in $\text{g/cm}^3$ and $E_0$ is in keV.
 On the other hand, theoretical considerations by Kanaya and Okayama \cite{kanaya1972penetration} lead to
\begin{align}
\label{eq:Kanaya}
 R= 2.76\times10^{-2}\frac{AE_0^{1.67}}{\rho Z^{0.89}}\;[\mu\text{m}],
\end{align}
where $Z$ is the atomic number, $A$ is the atomic mass and $E_0$ is in keV. The following composite
formula proposed by Fitting \cite{fitting1974transmission} has been used by several authors
\cite{fitting1976multiple,fitting2002cathodoluminescence,
meyza2003secondary,touzin2006electron,gaber1984energy}:
\begin{align}
\label{eq:FittingEarly}
R=
\begin{cases}
900 \rho^{-0.8}E_0^{1.3}\;[\mbox{\AA}] &\text{for} \; E_0<10\ \text{keV},
\\
450 \rho^{-0.9}E_0^{1.7}\;[\mbox{\AA}] &\text{for} \; E_0>10\ \text{keV},
\end{cases}
\end{align}
where $ \rho $ is in $\text{g/cm}^{3}$.
Here we employ the most recent estimate by Fitting \cite{fitting2011secondary}:
\begin{align}
\label{equ:depth}
R(\rho,E_0)=93.4\frac{E_0^{1.45}}{\rho^{0.91}}\;[\text{nm}],
\end{align}
where $\rho$ is in $\text{g/cm}^3$ and $E_0$ is in keV.

According to several studies\cite{fitting1977electron,cornet2008electron,touzin2006electron}
the actual distribution of the secondary electrons and holes
is well-approximated by a three-dimensional Gaussian function with its focus $\textbf{x}_0$ located $0.3R$
below the vacuum-sample interface:
  \begin{align}
  \label{eq:gFunction}
  g(\mathbf{x},E_0)=\frac{\alpha A}{E_{\rm i}} \exp\left(-B\vert\textbf{x}-\textbf{x}_0\vert^2\right),
  \end{align}
where $E_{\rm i}$ is the mean creation energy for one SE, $\alpha$ is the yield
factor close to one, and
 \begin{align}
 B=\frac{7.5}{R^2},\;\;\; A=\frac{BC}{\pi}.
 \end{align}
The constant $C(E_0)$ is proportional to the fraction $\eta$ of backscattered PE.
For silicon, silicon dioxide, and aluminium oxide, with $\eta\approx 0.2$,
$C$ can be obtained from:
\begin{align}
C=1.544\frac{E_0}{R},
\end{align}
where $C$ is in eV$\mbox{\AA}^{-1}$ and $E_0$ is in keV.

To account for the action of the surface potential $V_{\rm s}$ on the primary electron,
we introduce the effective energy $E_{\rm eff}=E_0 + V_{\rm s}(t_{i})$, where
$t_{i}$ is the time of impact, which should be applied in the
distribution function instead of $E_{0}$, thus arriving at:
\begin{align}\label{equ:Gus}
\begin{split}
g(\mathbf{x},E_{\rm eff})&=11.58\frac{E_{\rm eff}}{\pi R^3 E_{\rm i}}
\exp\left(-\frac{7.5}{R^2}\vert\textbf{x}-\textbf{x}_0\vert^2\right),
 \end{split}
\end{align}
where $R(\rho,E_{\rm eff})$ is given by (\ref{equ:depth}) and the pair creation energy
$E_{\rm i}$ depends on the material of the sample via\cite{touzin2006electron}
\begin{align}
E_{\rm i}\approx 3\,E_{\rm g}+1\,\text{eV},
\end{align}
with $E_{\rm g}$ denoting the energy gap of the material in eV.

The total numbers $N_{SE,SH}$ of secondary electrons and holes corresponding to
the distribution (\ref{equ:Gus}) can now be estimated as
\begin{align}\label{equ:num}
\begin{split}
N_{SE}&=N_{SH}
\approx\iiint\limits_{{\mathbb R}^{3}, z\geq 0}
g(\mathbf{x},E_{\rm eff})dV
\approx 0.877\frac{E_{\rm eff}}{E_{\rm i}},
\end{split}
\end{align}
showing that approximately $88\%$ of the effective energy is spent on the
creation of charge pairs, which generally agrees with MC simulations.
According to (\ref{equ:num}) the number of secondary electrons generated by one primary
electron is somewhere between hundreds and thousands. Hence, we may expect
the drift-diffusion-reaction approach to be a reasonable approximation at this scale.

Thus we take the $h_{p}$ for holes in (\ref{eq:SourceFunctionsFactors}) to be
equal to $g$ as introduced in (\ref{equ:Gus}).
Whereas for the electrons we recall that the primary electron
is still present in the sample at $t=t_{\rm g}$.
Hence, we adjust the coefficient
in front of the exponent in the function $g$ so that
it features one additional particle upon the integration (\ref{equ:num}).
Thus, the factors of eq.~(\ref{eq:SourceFunctionsFactors}) are
\begin{align}
\label{eq:SourceSpatialFactors}
\begin{split}
h_{n}(\mathbf{x},E_{\rm eff})&=
\left(11.58\frac{E_{\rm eff}}{E_{\rm i}}+13.158\right)\times
\\
&\frac{1}{\pi R^3}\exp\left(-\frac{7.5}{R^2}\vert\textbf{x}-\textbf{x}_0\vert^2\right),
\\
h_{p}(\mathbf{x},E_{\rm eff})&=g(\mathbf{x},E_{\rm eff}),
\end{split}
\end{align}
where $E_{\rm eff}$ depends on the surface electric potential at the time of impact.
Of course, the source functions proposed here are only approximations. Nevertheless,
they are based on the best experimental evidence and first principles calculations
available to date.

\subsection{Bombardment and temporal smoothing}
Depending on the beam current primary electrons may arrive at an average rate
as high as tens of millions per second.
Previous applications of the drift-diffusion-reaction approach typically describe the
SEM beam as a constant flux of electrons.
The goal of the present paper is to avoid the latter approximation and directly
consider, say, $m$, primary electrons arriving at times $t_{i}$, $i=0,1,\dots,m$.
Thus, one obtains a pulsed source where the next PE arrives in a medium with
some residual charge left from the impact of the previous PE.

Although, we gain some valuable insights about the subsurface charge dynamics and
the effect of beam current,
it is obviously too time consuming to consider bombardments of
a sample by a large number of electrons in this way. Hence, a different approach
is needed to study saturation effects at larger time scales.
Also, the SE yield calculations on the
level of single PE's, although possible, are hard to justify and interpret.

The main technical challenge preventing direct large-scale simulations with
our method is the pulsed nature of the source terms
requiring many time steps to be performed by the solver between electron impacts.
A way to reduce the computational burden is to derive a smoother
function describing the behavior of source terms at larger time scales.
In the limit such a smoother source function should approach the constant beam currents
of the other DDR models.

To achieve this we employ a temporal average of our source function,
which also mimics the way the SEM response is measured (time-averaged yield, rather
than the yield due to individual PE's).
The average value of $S_n(\textbf{x},t)$ over a
period of time $T$ between the impacts can be expressed as
\begin{align}
 \label{eq:UniformSource}
 \bar{S}_n(\mathbf{x})=\frac{1}{T}\int_{0}^{T}S_n(\mathbf{x},t')dt',
\end{align}
and is a time-independent function. In what follows we call this a
time-uniform or simply a uniform source.

Unfortunately, smoothing of the source has its price.
Due to the presence of nonlinear
terms in (\ref{equ:Po})--(\ref{equ:hole-dens}),
solutions obtained with a time-averaged source term will not be the
exact time-averaged values of the unknowns, but only the
approximations thereof.
Hence, to apply the DDR approach at both
time scales successfully one needs to define constitutive relations and
material parameters, such as the surface recombination velocity,
for each scale separately. This is the so-called homogenization problem,
typical for spatial multiscale analysis in physics (e.g. effective medium
problem in electrodynamics).

Further, although a uniform source switched on at $t=0$ may be expected
to eventually produce a steady-state distribution of charge, it is an open
theoretical question whether the actual pulsed source leads to the corresponding
periodic charge variations.

%%%%%%%%%%%%%%%%%%%%%%%%%%%%%%%%%%%%%%%%%%%%%%%%%%%%%%%%%%%%%%%%%%%%%%%%%%%%%%%%%%%%%%%%%%%%%%%%%%%%%%%%%%
%%%%%%%%%%%%%%%%%%%%%%%%%%%%%%%%%%%%%%%%%%%%%%%%%%%%%%%%%%%%%%%%%%%%%%%%%%%%%%%%%%%%%%%%%%%%%%%%%%%%%%%%%%%%%
\section{Numerical method}

\subsection{Numerical scaling}
To avoid numerical difficulties and maintain the accuracy of the solution,
a simple scaling of variables has been performed.
To this end we introduce a set of characteristic dimensionless quantities.
We denote the characteristic length scale by $l^*$, the characteristic
time scale by $t^*$ and the characteristic density scale by $\rho^*$.
The numerical values of these dimensionless parameters relevant
to the scale of the present problem are
\begin{align}
l^*=10^{-6},\ \ \ \ t^*=10^{-12},\ \ \ \ \rho^*=10^{18}.
\end{align}
There is a relation between these values ($t^*=(l^*)^2$ and $\rho^*=(l^*)^{-3}$)
that doesn't change the form of the equations, so that one only needs to introduce
the rescaled versions for some of the constitutive parameters:
 \begin{align}
   \begin{split}
 &\tilde{\tau}_{n,p}=\frac{\tau_{n,p}}{t^*},\ \ \tilde{n_i}=\frac{n_i}{\rho^*},\ \ \tilde{\varepsilon}=\frac{\varepsilon}{\rho^*(l^*)^2},\\
 &\tilde{\sigma}_{n,p}=t^*\rho^*\sigma_{n,p},\ \ \tilde{\gamma}_{n,p}=t^*\gamma_{n,p},\\
 &\tilde{N}_{n,p}=\frac{N_{n,p}}{\rho^*},\ \ \tilde{S}=\frac{t^*}{\rho ^*}S=\frac{dN}{d\tilde{t}}g(\tilde{\textbf{x}}),
   \end{split}
 \end{align}
Also the boundary and initial conditions should be rescaled, since, e.g.
the rescaled version of the surface recombination velocity is given by:
\begin{align}
 \tilde{v}_n=l^*v_n.
\end{align}

\subsection{FEM solver}
We employ the finite element method (FEM) for the numerical solution of
the coupled system (\ref{equ:Po})--(\ref{equ:hole-dens}) and implement it as a solver
within the COMSOL Multiphysics package. Although, there are many predefined modules and solvers
in COMSOL, none of them can be directly applied with the present problem.
The closest match is the semiconductor module. However, it is neither suited for
studying the two different domains defined above, i.e., $\Omega_1\cup\Omega_2$
for equation (\ref{equ:Po}) and $\Omega_2$ for the rest, nor does it
allow to incorporate the additional equations (\ref{equ:nTrap}) and (\ref{equ:pTrap}).
Therefore, we have opted for building a new model using the general PDE and
the ODE/DAE interfaces of COMSOL.

Since the charge densities may be extremely concentrated around the impact zone
and form very thin layers near the vacuum-sample interface, a careful discretization
strategy is required. To achieve sufficient accuracy one is advised to use the adaptive mesh with
refinement in the impact zone and at the interface as well as
the second-order Lagrange shape functions. A fully coupled approach with Newton-Raphson
solver and adaptive time-stepping algorithm has shown the best performance
with the present problem.

\begin{figure}
\centering
\includegraphics[scale=0.4] {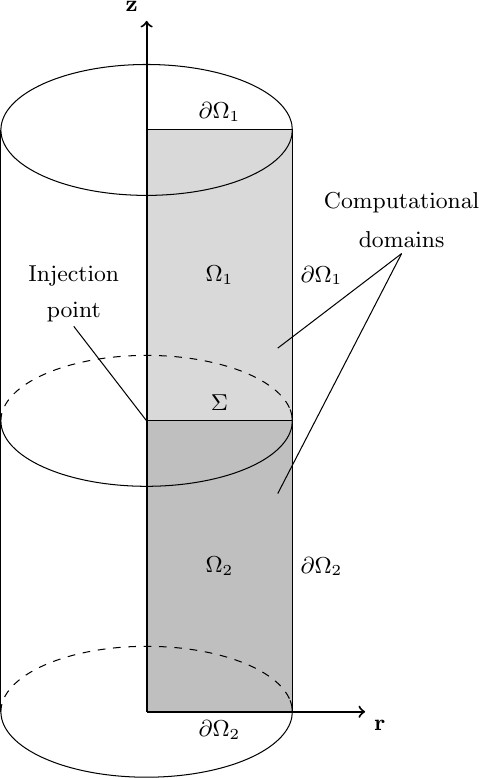}
\caption{Cylindrical geometry. }
\label{Raftari2015fig2}
\end{figure}

The computational complexity of the problem prohibits a full three-dimensional (3D)
simulation of realistic domains with sufficient spatial resolution on a standard PC.
Nevertheless, one can obtain 3D results for certain configurations at a typical
two-dimensional (2D) cost by exploiting their symmetry. Consider, for example,
the cylindrical geometry presented in Figure~\ref{Raftari2015fig2}.
In the cylindrical coordinate system $(r,\theta,z)$
the PE beam impinging along the $z$-axis corresponds to the source term
and boundary conditions independent of the angular coordinate $\theta$.
The solution will also be independent of $\theta$ and the original 3D model
is reduced to a 2D model in the $(r,z)$-coordinates.
To arrive at the corresponding FEM solver the PDE's
(\ref{equ:Po}), (\ref{equ:electron}) and (\ref{equ:hole}) must be written in the
so-called weak form, which is derived in the Appendix.

%%%%%%%%%%%%%%%%%%%%%%%%%%%%%%%%%%%%%%%%%%%%%%%%%%%%%%%%%%%%%%%%%%%%%%%%%%%%%%%%%%%%%%%%%%%%%%%%%%%%%%%%%%%%%
\section{Numerical experiments}

\subsection{Parameters and parameter-tuning}
It is clear that the values of the many constitutive parameters
in the equations (\ref{equ:Po})--(\ref{equ:hole-dens})
may have considerable influence on the results of simulations.
In the ideal situation these parameters are either measured
in dedicated experiments or computed from the first principles of quantum physics.
While this status-quo has long been established with the usual semiconductor
materials, the data for insulators are virtually absent.
Often, very different values for the same parameter are reported in
the literature, which could be the result of varying sample properties,
experimental conditions, or even trivial human errors (see e.g. detrapping rates in
\cite{hwang2009drift} and \cite{vasudevan1991numerical}).
In particular, there is a lot of uncertainty about the parameters of
recombination processes in the bulk and at interfaces.
With this in mind we have made a selection of typical values for two
distinct dielectrics shown in Table~\ref{tab:I}.

%%%%%%%%%%%%%%%%%%%%%%%%%%%%%%%%%%%%%%%%%%%%%%%%%%%%%%%%%%%%%%%%%%%%%%%%%%
\begin{table}[t]
\centering
\caption{Parameters of dielectric materials.}
    \begin{tabular}{|l|l|l|l|}
   \hline
    \small{Parameter}& $\text{SiO}_2$ & $\text{Al}_2\text{O}_3$& Unit\\
    \hline
    &&&\\
    $\varepsilon$ & $3.9$ (ref.\cite{kwo2001properties}) & $10$ (ref.\cite{meyza2003secondary}) & \\
    $\mu_n$ & $20$ (ref.\cite{renoud2004secondary})& $4$ (ref.\cite{hughes1979generation})& $\text{cm}^2\text{V}^{-1}\text{s}^{-1}$  \\
    $\mu_p$ & $0.01$ (ref.\cite{renoud2004secondary})& $0.002^*$   &$\text{cm}^2\text{V}^{-1}\text{s}^{-1}$  \\
    $\sigma_n$& $10^{-15}$ (ref.\cite{vasudevan1991numerical})& $10^{-15}$  & cm$^2$ \\
    $\sigma_p$ & $10^{-18}$ (ref.\cite{vasudevan1991numerical})& $10^{-18}$ & cm$^2$  \\
    $v_{th}$ & $10^7$ (ref.\cite{renoud2004secondary})&  $10^7$ &cm s$^{-1}$  \\
    $\tau_{n,p}$ & $2\times10^{-9}$ (ref.\cite{hughes1979generation}) & $2\times10^{-9}$  & s  \\
    $\rho$ & $2.65$ (ref.\cite{helms1994silicon}) & $3.98$ (ref.\cite{meyza2003secondary}) &  g cm$^{-3}$  \\
    $E_{\rm g}$ & $9$ (ref.\cite{meyza2003secondary}) & $9$ &eV  \\
    $N_{n,p}$  & $1.6\times10^{19}$ (ref.\cite{renoud2004secondary})&  $1.6\times10^{19}$ & cm$^{-3}$\\
    $\gamma_{n,p}$ & $10^4$ (ref.\cite{hwang2009drift}) & $10^4$ & s$^{-1}$ \\
    $k$ & 25 & 25&s$^{-1}$\\
    $w$ & $10^{-5}$ &$10^{-5}$ &\\
    \hline
    \end{tabular}

    *{\small This value could not be found in literature and has been chosen
    by analogy with the relation between
    the electron and hole mobilities in $\text{SiO}_2$.}
\label{tab:I}
\end{table}

In the boundary condition (\ref{equ:interface1}) the surface recombination velocity
$v_{n}$ plays an important role analogous to the reflection coefficient of the
current-based approach.
In the literature the SRV is mostly discussed and
measured for the metal-oxide-semiconductor interfaces.
With respect to our case, which is a vacuum-semiconductor interface,
the only reference that could be found refers to
Germanium under ion bombardment\cite{PhysRev.112.793}, reporting
the value of $(5\ \text{to}\ 7)\times 10^3\ \text{cm}/\text{s}$.
Considering the SRV to be material-dependent we believe that its value
should be determined based on the consistency between
the results of the present model and dedicated experiments.

The SE yield is defined as the number of secondary electrons emitted
through the sample-vacuum interface and picked up by the detector per one
incident electron. The importance of the SE yield stems from the fact that it
is one of the few directly measurable quantities in SEM\cite{Database}.
Monte Carlo simulations of the SE yield are also available
\cite{renoud2004secondary,renoud2002monte,kieft2008refinement,ding2001monte}.
Therefore, in the tuning of the SRV parameter one would mostly be relying
on the SE yield data as a function of the PE energy. In the present
continuous approximation the SE yield is computed as
the flux density through the boundary $\Sigma$ integrated over
this boundary and over time from $t=0$ to $t=t_{\rm end}$,
and divided by the number of PE's that arrived at the sample during
that time interval.

The following steps describe a simple optimization procedure
for tuning the value of SRV:
\begin{itemize}
\item Let $Y_{\rm exp}$ be the SE yield measured fro PE's with energy $E_{0}$.
\item Let $v_{n}^{(0)}$ be the initial guess for the SRV,
and let $Y(v_{n}^{(0)})$ be the SE yield computed
by the DDR solver with the SRV set to $v_{n}^{(0)}$.
\item For $v_{n}^{(0)}$ sufficiently close
to the true (optimal) value we can assume a linear relation:
\begin{align}
 Y_{\rm exp}-Y(v_n^{(0)})=\alpha(v_n-v_n^{(0)}).
\end{align}
Since, obviously, $Y(0)=0$, the coefficient $\alpha$ can
be obtained as $\alpha=\frac{Y(v_n^{(0)})}{v_n^{(0)}}$,
so that $v_n^{(1)}=v_n^{(0)} \frac{Y_{\rm exp}}{Y(v_n^{(0)})}$.
\item{Compute $Y(v_n^{(1)})$ with the DDR solver.}
\item If $Y(v_n^{(1)})$ is sufficiently close to $Y_{\rm exp}$,
then stop and set $v_{n}=v_n^{(1)}$.
Otherwise, continue with $v_n^{(1)}$ as the new initial guess.
\end{itemize}
In principle, this process should be repeated with the SE yield data
for a whole range of PE energies $E_{0}$.
Unless changes in $E_{0}$ significantly alter the temperature of the
sample, the SRV of a given material is supposed to be independent of $E_{0}$.
Hence, if the corresponding tuned
values of $v_{n}$ for some material are all close to each other,
then we have an additional confirmation that the DDR method
is working properly.

We conclude this section with a word of caution concerning the
use of SE yield in determining the SRV. First of all, typical SE
yield data correspond to some kind of stationary regime.
It is known, however, that the SE yield keeps changing after the start of
bombardment for quite a long time. Hence, we can compare the results of
simulations with experimental data and tune the $v_{n}$ parameter only
upon bombardment of the sample with a sufficiently large number of PE's. In our approach
such long-time simulations can only be performed with the time-uniform source.
This means, however, that the tuned value of $v_{n}$ will be effective
in nature.

Secondly, numerical experiments demonstrate that the electron flux
through the sample-vacuum interface
is not only time-dependent, but also depends on the extent of the sample
and the proximity of ohmic contacts or other conducting materials to the
beam's entry point. Hence, one may expect different SE yield values
with different samples of the same material.

\subsection{Impact of a single primary electron\label{sec:SingleElectron}}
In this section we investigate the events following the
injection of a single primary electron into a neutral dielectric sample.
The goal of these numerical experiments is to estimate the space-time scales of the
dynamics separately for all four particle species, i.e., $n$, $p$, $n_{t}$, and $p_{t}$,
as well as the total charge density $q(p+p_{t}-n-n_{t})$ and the potential $V$.
In particular, these simulations will help us to demonstrate that despite
the poor mobility and diffusivity of dielectrics,
the drift and diffusion of free charges is generally much faster than the
characteristic time scales of the charging process.

We focus on two common materials featured in many of the previous
studies, namely, on the oxides
$\text{SiO}_2$ and $\text{Al}_2\text{O}_3$.
As one can see from the data of Table~\ref{tab:I} the difference between
these materials is in the values
of the relative permittivity $\varepsilon$ ($\text{SiO}_2$ has a smaller
$\varepsilon$), the electron and hole
mobilities $\mu_{n,p}$ ($\text{SiO}_2$ is relatively
more conductive), and the mass density $\rho$ ($\text{SiO}_2$ has a smaller density).
The effective values of the SRV, $v_{n}=100$~cm/s for
$\text{SiO}_2$ and $v_{n}=200$~cm/s for $\text{Al}_2\text{O}_3$, were
obtained with the optimization procedure explained above and typical
experimental SE yield data \cite{barnardmeasurements,yong1998determination,Database} using the
time-uniform source model. Below we focus on a fixed PE energy $E_{0}=1$~keV.

\begin{figure}
\includegraphics[scale=0.28] {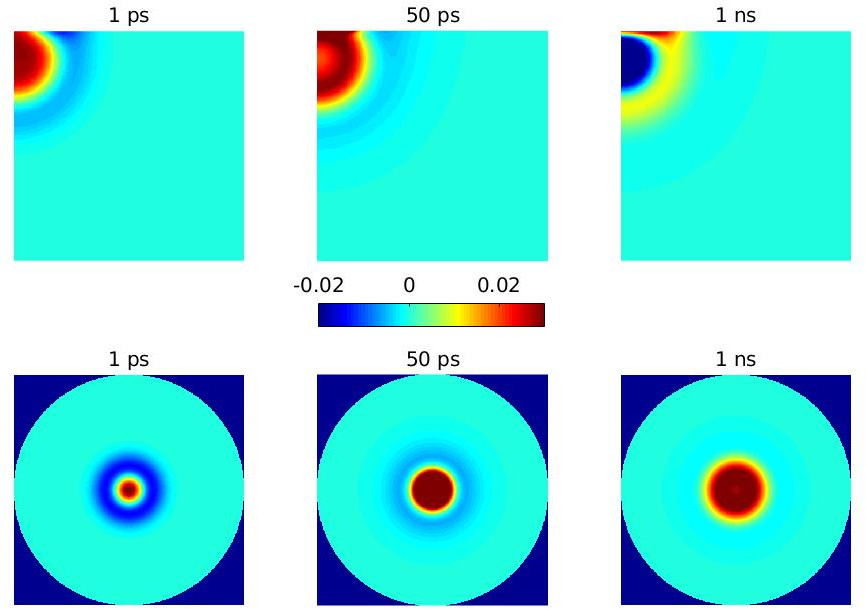}
\caption{Evolution of the total charge $(p+p_t-n-n_t)q$ (C$\,\text{cm}^{-3}$)
in $\text{SiO}_2$
after the impact of a single primary electron with the energy $\text{E}_0=1$ keV.
Top row: vertical cross-section, side length is $100$~nm; bottom row: 
top-view of the sample-vacuum interface, diameter is $200$~nm.
}
\label{Raftari2015fig3}
\end{figure}

\begin{figure}
\includegraphics[scale=0.3] {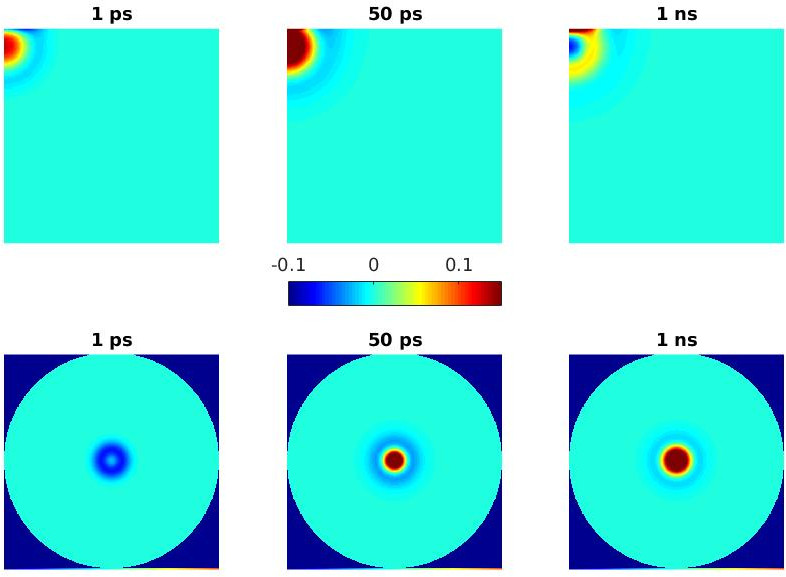}
\caption{Evolution of the total charge $(p+p_t-n-n_t)q$ (C$\,\text{cm}^{-3}$)
in $\text{Al}_2\text{O}_3$
after the impact of a single primary electron with the energy $\text{E}_0=1$ keV.
Top row: vertical cross-section, side length is $100$~nm; bottom row: 
top-view of the sample-vacuum interface, diameter is $200$~nm.
}
\label{Raftari2015fig4}
\end{figure}

Figures \ref{Raftari2015fig3} and \ref{Raftari2015fig4} show the snapshots of the time evolution of
the four charge species in the two materials.
As can be seen from the upper-left images of these figures,
the smaller mass density of $\text{SiO}_2$ means,
see eq.~(\ref{equ:depth}),
that with the same $E_{0}$ the maximum PE penetration depth $R$ and the center
of the initial charge distribution are deeper for $\text{SiO}_2$
than for $\text{Al}_2\text{O}_3$. The overall shapes of the initial charge
distribution are different as well, see eq's.~(\ref{equ:Gus}) and
(\ref{eq:SourceSpatialFactors}), with the one of $\text{SiO}_2$ being
broader. Hence the DDR dynamics starts with different initial states
in these materials.

The generation of charge pairs by ionization takes place in the period of 1 picosecond after injection
($t_{\text{g}}$=1 ps). Note that in our model the processes of recombination and trapping begin
already at $t=0$.
At $t_{\text{g}}$ the density of free electrons is already beginning to decrease. In fact,
the density of free electrons reaches its maximum of $2.07\times10^{18}\ \text{cm}^{-3}$ and
$6.35\times10^{18}\ \text{cm}^{-3}$ for $\text{SiO}_2$ and $\text{Al}_2\text{O}_3$, respectively,
at around $t=0.6$~ps.
The density of free holes reaches its maximum roughly at $t=0.7$~ps and
remains constant till the end of the generation period $t_{\text{g}}$.
The maximum density of free holes in this stage for $\text{SiO}_2$ is $2.31\times10^{18}\ \text{cm}^{-3}$ and for $\text{Al}_2\text{O}_3$
is $7.01\times10^{18}\ \text{cm}^{-3}$.

As expected, the density of trapped electrons and holes initially increases with time
reaching, respectively, (at $t_{\text{g}}$) the values of $1.79\times10^{17}\ \text{cm}^{-3}$
and $2.02\times10^{14}\ \text{cm}^{-3}$ in $\text{SiO}_2$, and
$5.31\times10^{17}\ \text{cm}^{-3}$ and $6.09\times10^{14}\ \text{cm}^{-3}$ in $\text{Al}_2\text{O}_3$.
The lower density for the trapped holes compared with trapped electrons is due to the
smaller trapping cross sections.
In $\text{SiO}_2$ the density of trapped electrons reaches its maximum
of $2.01\times10^{18} \ \text{cm}^{-3}$ at $t=50$~ps.
Then, for more that $50$~ns, which is a relatively long time,
no change is seen in the distribution of trapped electrons. After that
the maximum density of trapped electrons starts to decrease
dropping to $1.69\times10^{18}\ \text{cm}^{-3}$ at $t=1$~$\mu$s.
For the trapped holes, reaching the maximum density takes much longer time
compared to the trapped electrons.
The density of trapped holes keeps increasing at $t=1$~$\mu$s
and reaches its maximum of $1.71\times10^{18}\ \text{cm}^{-3}$ at $t=50$~ns.
The maximum density for trapped holes at time $t=1$~ns is $2.74\times10^{17}\ \text{cm}^{-3}$
and at $t=1$~$\mu$s the density of the trapped holes is $1.7\times10^{18}\ \text{cm}^{-3}$.
A similar dynamics of trapped particles is observed in $\text{Al}_2\text{O}_3$.

During the first microsecond a fast decrease in the density of free electrons
and a slower decrease in the density of free holes is observed owing to the
relatively strong trapping of electrons and a weaker trapping of holes.
In $\text{SiO}_2$ the major drop in the density of free electrons happens during the first $50$~ps.
The maximum density of free electrons is $5.21\times10^{14}\ \text{cm}^{-3}$ at $t=50$~ps
and reaches almost the intrinsic carrier density of the material at $t=1$~$\mu$s.
At $t=50$~ps the density of free holes is
higher than that of the free electrons ($2.11\times10^{18}\ \text{cm}^{-3}$).

The interplay of the four charge species leads to the total charge
density resembling an expanding spherical wave with initially
a positive charge region in the middle surrounded by a shell of negative charge followed
by a negative middle region with a positive shell.
Due to the emission of electrons the initial predominantly negative charge at the sample-vacuum
interface is gradually replaced by the positive charge.
In $\text{SiO}_2$, the positive charge reaches its maximum of $0.13$~C/cm$^3$ at the surface at $t=50$~ps and the
negative charge has the maximum of (in the sens of absolute value) $-0.05$~C/cm$^3$ at $t=2$~ns and is situated close to the surface .
In $\text{Al}_2\text{O}_3$, the positive charge increases its maximum of $0.4$~C/cm$^3$ at the surface at time $t=200$~ps
and the negative charge has the maximum of $-0.23$~C/cm$^3$ in the center of impact zone at time $t=18$~ns.

Obviously, the electric potential closely follows the distribution of the total charge.
Initially we observe a positive potential in the middle of the impact zone surrounded
by a shell of weak negative potential.
For example, in $\text{SiO}_2$ at the beginning a positive potential with the maximum of
$0.11$~V is observed stretching across the sample-vacuum interface.
A shell of weak negative potential is situated around this
positive central region and appears inside the sample only.
After a few hundred picoseconds, a transition occurs
which results in a different situation for potential and that is the negative potential appears
in the middle with shells of positive and negative potentials, respectively.
The minimal value achieved by the potential during the
first microsecond is $-0.1$~V in $\text{SiO}_2$ which is situated in the center of impact zone and in the
period of $20$ to $50$~ns.

\subsection{Electron bombardment}

\begin{figure}
\includegraphics[scale=0.29] {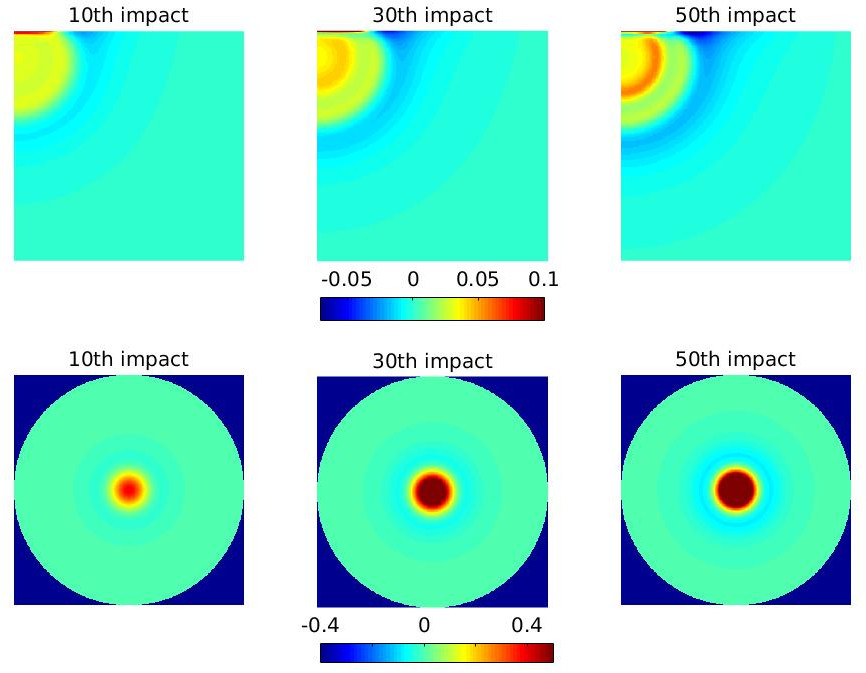}
\caption{Evolution of the total charge $(p+p_t-n-n_t)q$ (C$\,\text{cm}^{-3}$)
in $\text{SiO}_2$ during bombardment with a beam current of 160 nA.
Top row: vertical cross-section, side length is $100$~nm; bottom row: 
top-view of the sample-vacuum interface, diameter is $200$~nm.
}
\label{Raftari2015fig5}
\end{figure}

\begin{figure}
\includegraphics[scale=0.29] {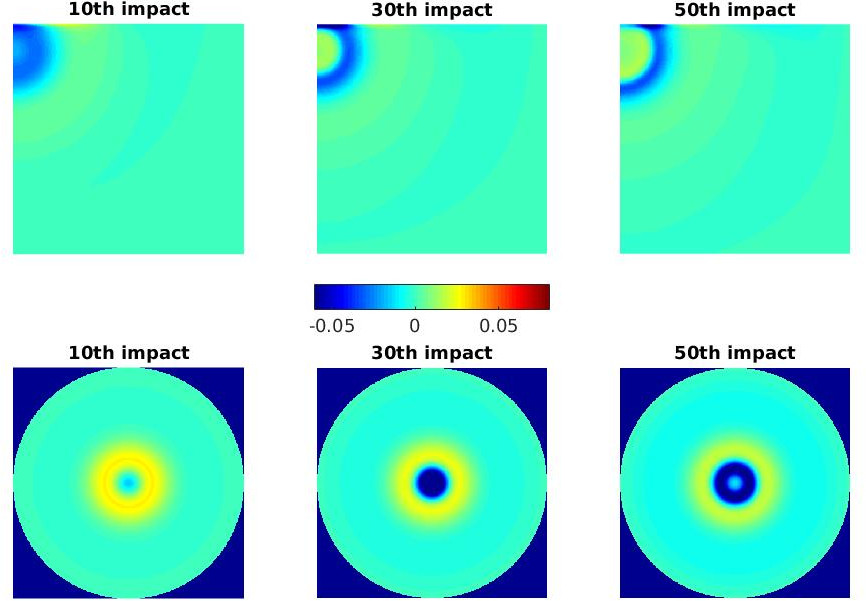}
\caption{Evolution of the total charge $(p+p_t-n-n_t)q$ (C$\,\text{cm}^{-3}$)
in $\text{SiO}_2$ during bombardment with a beam current of 160 pA.
Top row: vertical cross-section, side length is $100$~nm; bottom row: 
top-view of the sample-vacuum interface, diameter is $200$~nm.
}
\label{Raftari2015fig6}
\end{figure}

The electron gun of a typical SEM is able to produce PE currents in the range
of pico to nano Amp{\'e}res (i.e. average interval between PE impacts from
nano to picoseconds).
The charge dynamics following the impact of a single PE, analyzed in the previous section,
clearly shows that the next electron faces highly variable
conditions in the sample depending on the time of its arrival.

Since the main features of the charge dynamics in $\text{Al}_2\text{O}_3$ and $\text{SiO}_2$ are
essentially similar, we restrict our discussion to the latter material.
In this section a $\text{SiO}_2$ sample is considered under focused beams
with the currents of 160 nA and 160 pA (average times between PE impacts are
$1$~ps and $1$~ns, respectively).
To illustrate the nontrivial effect of the varying PE current the results in Figures~\ref{Raftari2015fig5}
and \ref{Raftari2015fig6} are presented for the same number
of PE impacts in both beams that, obviously, correspond to different illumination times.

We start with the higher current of 160 nA modelled as a sequence of PE's arriving
with exact one picosecond intervals between them. Figure~\ref{Raftari2015fig5} shows the
evolution of the total charge density during the first 50 impacts.
On the fine temporal scale (not shown) we observe that
the densities of free electrons and holes reach their maxima of
$9.05\times10^{19}$ and $1.06\times10^{20}\ \text{cm}^{-3}$, respectively,
at the end of the generation (ionization) stage after impacts and decrease afterwards.

The maximum density of trapped electrons reaches its maximum (the density of trapping sites $N_{n}$)
for the first time at $t=30$~ps (after 30 PE impacts) inside the impact zone.
A similar local saturation for the trapped holes does not happen during the first 50 PE's, although,
their density grows.

Similar to the aftermath of a single PE impact we observe a (semi) spherical wave of charge density
emerging from the impact zone. However, now it remains a growing positive charge
zone surrounded by the shell of negative charge without the charge-sign oscillation as
in Figure~\ref{Raftari2015fig3}.
At the very beginning the positive charge has access to the surface, but towards the
$50$th impact a layer of negative charge prevents the positive charge from touching
the surface. The maximum positive charge of $1.36$ C/cm$^3$ is observed at
the end and at the surface (lower-left image of Figure~\ref{Raftari2015fig5}).
The positive charge in the center of the expanding zone
increases to $0.05$ C/cm$^3$ towards the end (upper-right image of Figure~\ref{Raftari2015fig5}).
The negative charge remains confined to a shell around the positive charge.
This shell becomes distorted by growing thicker with time along the sample-vacuum
interface with the distance from the injection point, thus, reaching the maximum
value of $-0.08$~C/cm$^3$ at the $50$th impact. 
In the beginning and at time $8$~ns, the postive potential reaches its maximum of $0.11$~V in the center of impact zone 
and the negative potential reaches the
minimum of $-0.15$~V at the end of this initial bombardment period.

Next, we consider the beam current of 160 pA corresponding to one nanosecond intervals between
PE impacts. Again, for a very short time after each impact,
an increase in the density of free electrons is observed, which, after less that $0.5$~ns,
drops to the almost the intrinsic carrier density
of the material. The maximum of $2.08\times10^{18}\ \text{cm}^{-3}$ occurs at the end of
the generation (ionization) stage.
The density of free holes reaches its maximum of $4.18\times10^{18}\ \text{cm}^{-3}$
in the middle of the generation stage.

For about $30$~ps after each impact an increase in the density of the trapped electrons
is seen after which the density remains constant until the next impact.
Comparing this with the previously considered higher current we observe that now
it takes a longer time ($21$~ns) but less impacts ($21$ PE impacts) for the
density of the trapped electrons to reach the density of trapping sites in the middle of the impact zone.
Similarly to the previous higher current, the density of trapped holes
does not reach the trapping site density during the considered bombardment period.

Comparing the surface charge of the high (Figure~\ref{Raftari2015fig5}, bottom) and
low (Figure~\ref{Raftari2015fig6}, bottom) currents
we see that with the higher current, the positive charge at the injection point is present surrounded by a ring of negative charge, 
whereas with the lower current the charge at the injection point is negative surrounded by a
ring of positive charge (Figure~\ref{Raftari2015fig5} bottom).
In fact, with the $160$~pA current the positive surface charge is also seen, but not at the time of impact.
Subsequently, after each impact, the positive potential is observed at the center of impact zone 
and after a few hundred picoseconds it is replaces with a negative potential. In the middle of the ionization time, both positive and negative 
potentials reach thier maxima of 0.17 V and -0.07 V, respectively.

\begin{figure}
\includegraphics[scale=0.27] {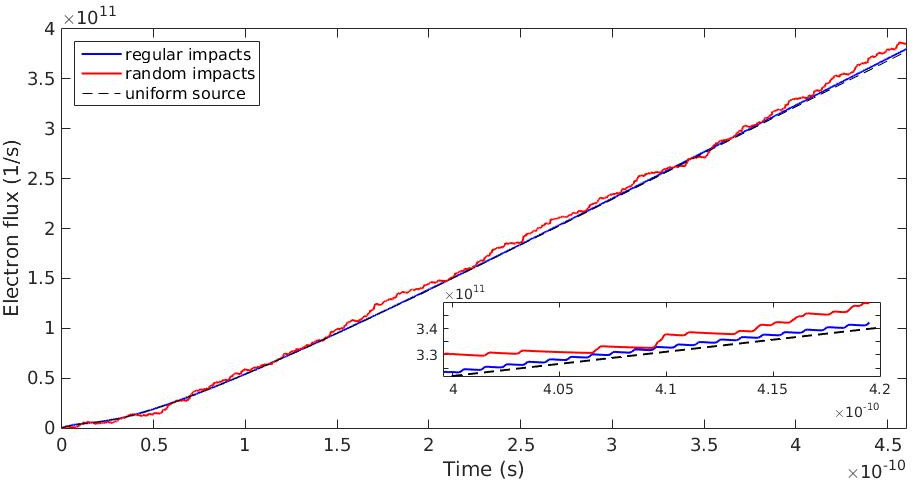}
\caption{Electron flux through the sample-vacuum interface in $\text{SiO}_2$ illuminated by the
160 nA beam obtained with the pulsed model (regular and random impacts) and the time-uniform source model.}
\label{Raftari2015fig7}
\end{figure}

Apart from the absence of rapid charge oscillation in the expanding spherical wave pattern, one
can notice another substantial difference with the dynamics of Figure~\ref{Raftari2015fig3}.
Namely, the visible speed of expansion of the charged zone is much slower under the
bombardment conditions, than during a single impact.
Later we shall see that this speed of expansion roughly
corresponds to the growth of the zone occupied by the trapped charges.

We have extended the simulation to $500$ PE impacts (at $160$~nA) and
compared the idealistic source with periodic impacts considered above
with the more realistic source whose PE's impact the sample at time instants
drawn from the Poisson distribution.
Another purpose of this $500$-impact simulation is to test the
applicability of the time-uniform source model, see eq.~(\ref{eq:UniformSource}).
Figure \ref{Raftari2015fig7} shows the electron flux through the
sample-vacuum interface obtained by the pulsed model with
regular and random impacts as well as time-uniform source model.
The result shows that for this relatively short period at the beginning of the bombardment
all three fluxes are in good agreement with each other.
Based on these promising results,
simulations for longer intervals of time can be carried out with the
time-uniform source model.

\subsection{Steady state}

From the mathematical point view no steady-state solution exists with
the pulsed source model (where each PE impact is modelled individually).
At most, one can expect a time-periodic solution if PE impacts happen at
regular intervals.
The
time-uniform source may, on the other hand, result in a solution that
converges to a steady state for $t\rightarrow\infty$. In this section
we investigate the large $t$ behaviour of the time-uniform source model
in various circumstances.

\begin{figure}
\includegraphics[scale=0.28] {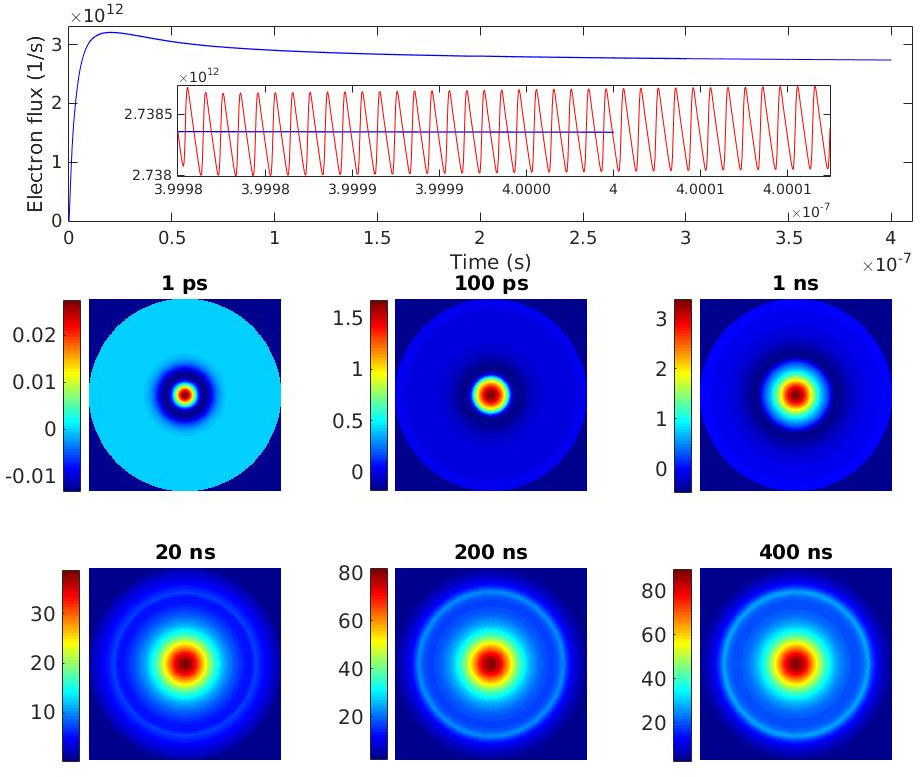}
\caption{Electron flux through the sample-vacuum interface (top plot).
Insert shows the zoomed-in portion of the plot corresponding to the end of the simulated
interval, where the oscillating curve represents flux obtained with the pulsed-source model.
Lower images: the evolution of the surface charge
obtained by the time-uniform source model with the $160$~nA beam current.}
\label{Raftari2015fig8}
\end{figure}

\begin{figure}
\includegraphics[scale=0.3] {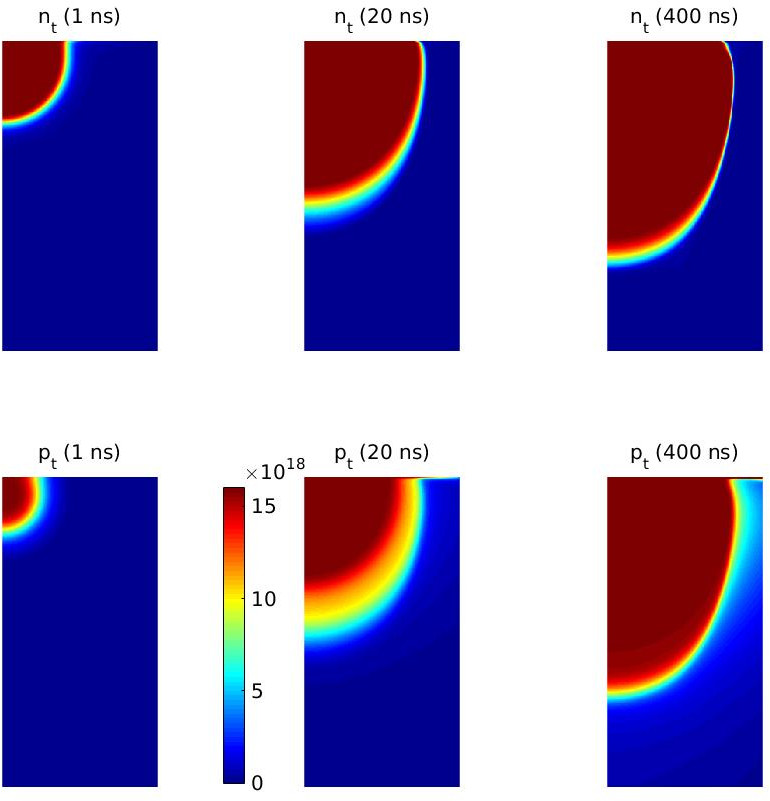}
\caption{The evolution of the densities of trapped electrons and holes obtained by
the time-uniform source model with the $160$~nA beam current. Vertical cross-section, width -- $100$~nm,
height -- $200$~nm}
\label{Raftari2015fig9}
\end{figure}

\begin{figure}
\includegraphics[scale=0.3] {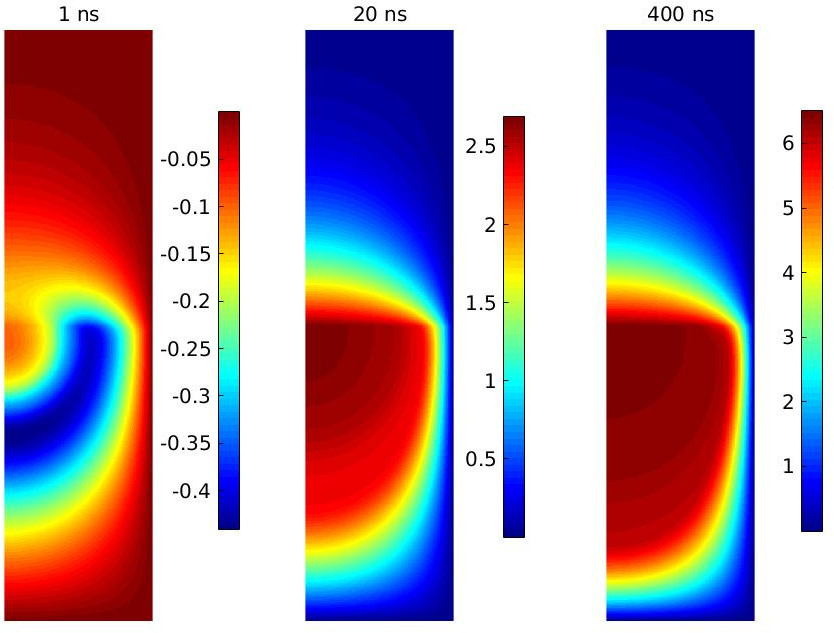}
\caption{The evolution of potential obtained by the time-uniform source model
with the $160$~nA beam current. Vertical cross-section (vacuum part included), width -- $100$~nm,
height -- $400$~nm}
\label{Raftari2015fig10}
\end{figure}

\begin{figure}
\includegraphics[scale=0.26] {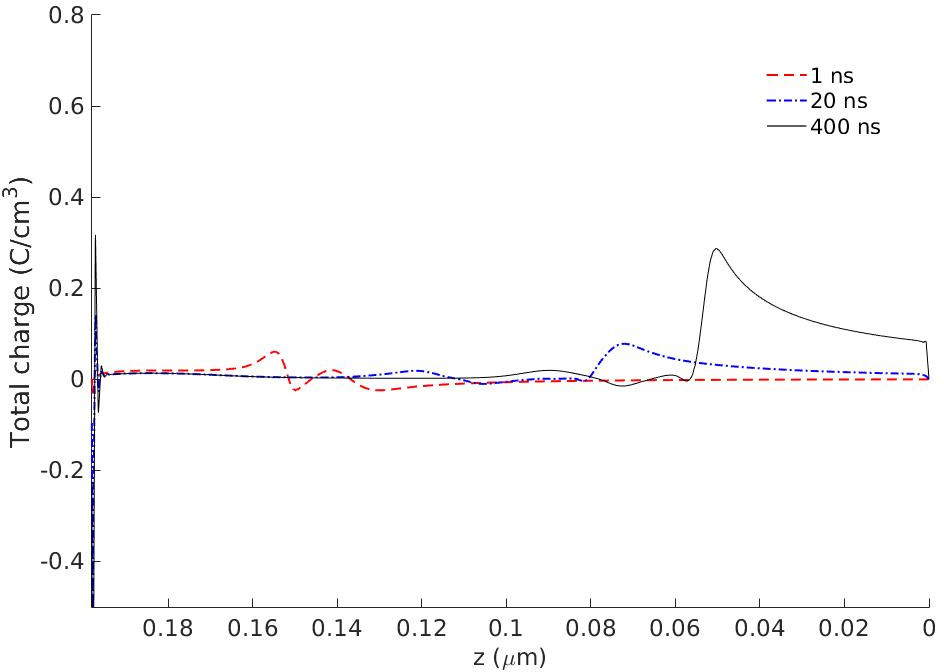}
\caption{Evolution of the total charge along the beam direction
($z$-axis) with the $160$~nA beam current.
The coordinate $z=0.2$ corresponds to the sample-vacuum interface. 
Vertical axis is clipped. See Fig.~\ref{Raftari2015fig8} for the actual surface charge values.}
\label{Raftari2015fig11}
\end{figure}

Figures~\ref{Raftari2015fig8}--\ref{Raftari2015fig11} show the simulation results
by the time-uniform source model in the case of the high
beam current of 160 nA. The electron flux through the sample-vacuum interface in
Fig.~\ref{Raftari2015fig8} shows that a steady state
starts around 400 nanoseconds in this case. To confirm that this, indeed, is
a steady state we initiate the pulsed-source bombardment of the sample with the
initial conditions set to the time-uniform steady-state solution.
If such a steady state is stable,
then the corresponding charge distributions and potential could be in
the neighbourhood of a time-periodic solution
expected in the case of the pulsed-source model with impacts at regular intervals.
As can be seen from the
magnified portion of the flux plot in Fig.~\ref{Raftari2015fig8}, the flux computed
with the pulsed-source model, indeed, appears to oscillate around the
time-uniform flux.

The evolution of the surface charge shown in Fig.~\ref{Raftari2015fig8}
starts with the positive charge at the injection point surrounded by a ring of negative charge. 
However, as the images show, after a few nanoseconds, this
negative charge ring is removed by an outward-going (surface) wave 
and the positive charge settles in the center as a steady state. This positive charge grows
up to the value of $88$~C/$\text{cm}^{3}$.

This behaviour and other features of the total charge dynamics
are strongly influenced by the evolution of the trapped charge density
shown in Fig. \ref{Raftari2015fig9}.
As one can see, the trapped charges rapidly reach their saturation value $N_{n}$ in the
impact zone, after which this saturated trapped-charge zone spreads outwards towards the
ohmic boundaries. The electron trapping process and the spread of the corresponding zone
is slightly faster than that of the holes (due to the higher mobility and trapping
cross-section of the electrons). This, in particular, explains the presence of a
slowly spreading negative shell in the top images of Figures~\ref{Raftari2015fig5}--\ref{Raftari2015fig6}.
Also, around $t=100$~ns, a surface channel reaching the omic contacts developes, 
consisting of saturated trapped holes.
It provides a path free of trapping for the positive charge transport along the
sample vacuum interface. The latter can explain the saturation of the 
positive surface charge density shown in Fig.~\ref{Raftari2015fig8}.

Further, the steady state with this beam current is characterized by the densities of free electrons and holes
around $2.31\times10^{21}\ \text{cm}^{-3}$ and $2.49\times10^{21}\ \text{cm}^{-3}$, respectively.
The evolution of potential is shown in Fig. \ref{Raftari2015fig10}.
It can be noted that potential follows the surface charge behavior, starting as
a negative ring around a positive impact region. Then, the negative potential moves to the ohmic boundaries
with time, and, after a few nanoseconds, the positive potential dominates in the sample
as well as the vacuum close to the interface.
A weak negative potential is observed below the positive one, however, it disappears
after a few nanosecond.

Since $N_{n}=N_{p}$ in the present simulations, and both the trapped electrons and the trapped holes
reach their saturation values almost everywhere (thus, effectively neutralizing the total trapped
charge), the positive potential at large $t$ is the result of free rather than trapped holes.
For example at $t=400$~ns, the free charge, $(p-n)q$, at the center of the surface is about
$89.4\ \text{C/cm}^{3}$ while the trapped charge, $(p_t-n_t)q$, at the surface 
and around the injection point is almost homogeneous and
around zero and close to the ohmic contact is $2.56\ \text{C/cm}^{3}$.

The evolution of the total charge along the beam direction when the beam current is $160$~nA is shown
in Fig. \ref{Raftari2015fig11}, where the charge is displayed at three different time instants. 
The slight spatial advance of the saturated trapped-electron zone with respect to
the saturated trapped-holes zone may explain the slowly expanding negative charge shell.
As mentioned above, in the equilibrium state the sample is positively charged by the free
holes. At this stage the trapped electrons and holes cancel each other, whereas,
the free electrons disappear not only at the ohmic contacts, but through the sample-vacuum
interface as well.

\begin{figure}
\includegraphics[scale=0.3] {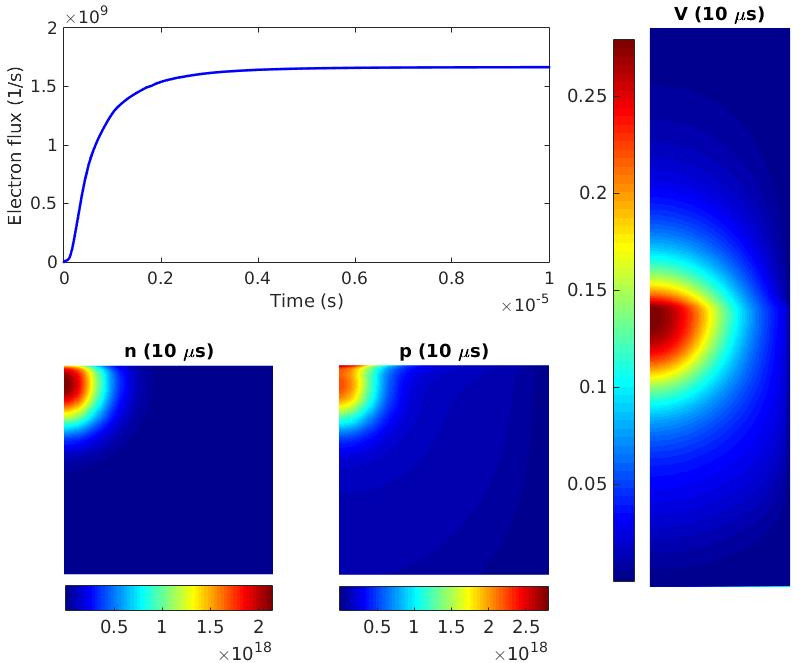}
\caption{Electron flux through the sample-vacuum interface, electric potential (vertical cross-section, vacuum part included, width -- $100$~nm,
height -- $400$~nm), and the densities of free charges (vertical cross-section, side length -- $100$~nm)
obtained by the time-uniform source model with the $160$~pA beam current.}
\label{Raftari2015fig12}
\end{figure}

Figure \ref{Raftari2015fig12} shows the simulation results with the lower $160$~pA beam current.
The electron flux through the sample-vacuum interface demonstrates that the system reaches
the steady state at a later time, compared to the previous higher $160$~nA current,
but with fewer PE impacts.
A significant difference is observed between the free charge distributions for these
two values of beam currents.
In particular, the shapes of the free electrons and free holes distributions
in the steady state are same in the case of higher $160$~nA beam current.
Whereas, the difference between these distributions at $160$~pA is obvious
from the images of Fig. \ref{Raftari2015fig12}.
The maximum density of free electrons is at the center of the impact zone,
while for the free holes it is at the surface.
The overall positive potential in the steady state
is the result of the excess of free holes at the surface.

\begin{figure}
\includegraphics[scale=0.28] {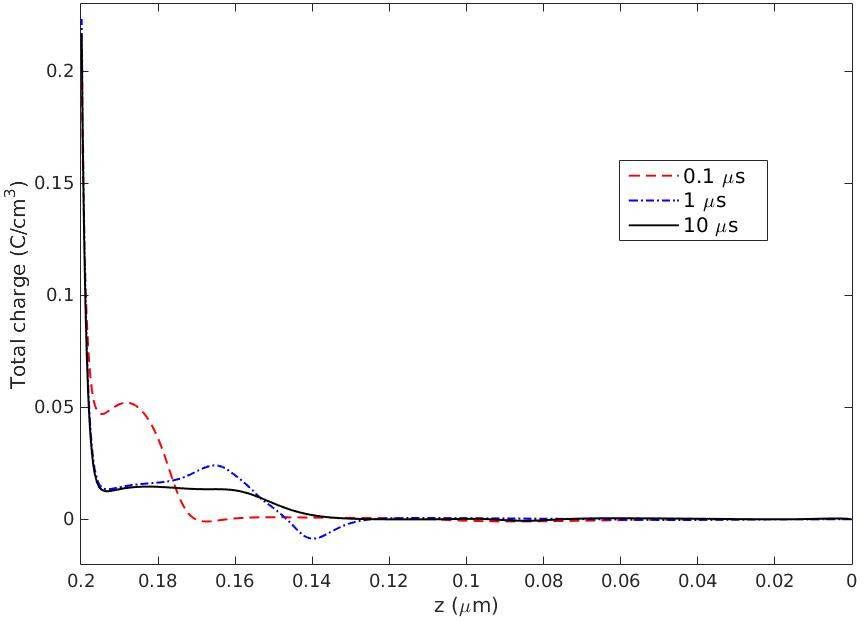}
\caption{
Evolution of the total charge along the beam direction
($z$-axis) with the $160$~pA beam current.
The coordinate $z=0.2$ corresponds to the sample-vacuum interface.}
\label{Raftari2015fig13}
\end{figure}

The total charge distribution along the beam direction for the beam current of $160$~pA
is shown in Fig.~\ref{Raftari2015fig13}.
It is apparent that the spatial variations in the total charge density happen closer
to the sample-vacuum interface (at depths less than $70$~nm) if compared to the
higher $160$~nA beam current.
Also, the surface charge remains positive from the start of the process.

\begin{figure}
\includegraphics[scale=0.285] {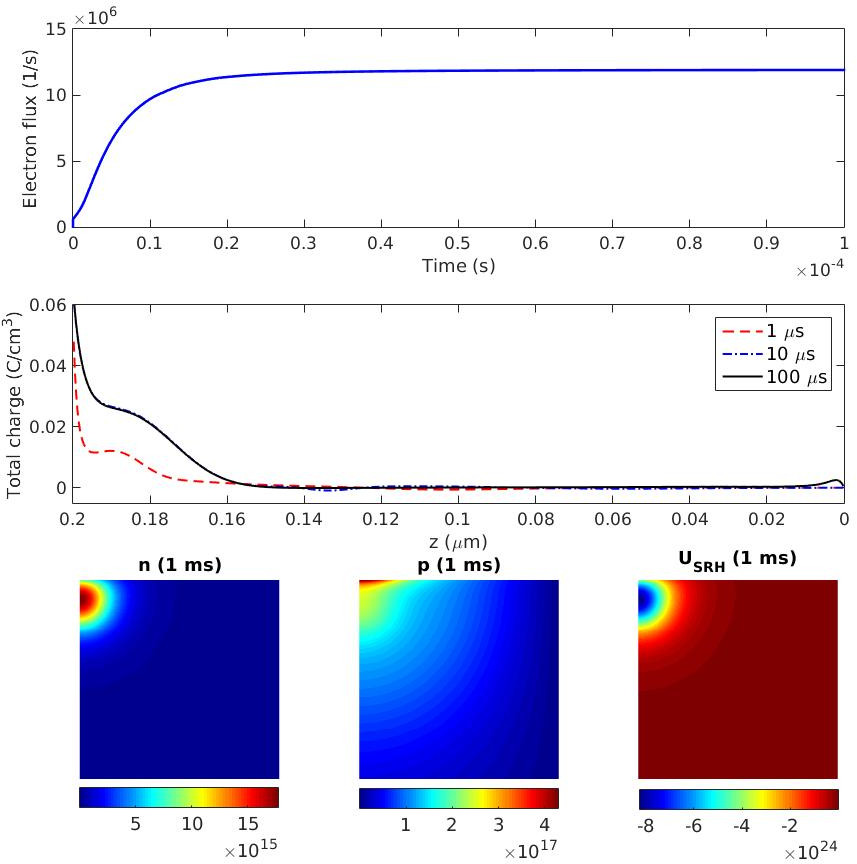}
\caption{Electron flux through the sample-vacuum interface and the evolution of
the total charge along the beam direction ($z$-axis) with the beam current of $16$~pA.
The bottom row shows the densities of free charges and the recombination rate
in the steady-sate regime (vertical cross-section, side length -- $100$~nm).}
\label{Raftari2015fig14}
\end{figure}

Finally, we consider an even lower $16$~pA beam current.
The electron flux through the sample-vacuum interface in Fig. \ref{Raftari2015fig14} and comparison with the
previously obtained fluxes due to higher beam currents bring us to the conclusion
that the flux rate and the time it takes to reach the steady state are roughly proportional to the
beam current. Yet the number of PE impacts to reach the steady state is roughly
inversely proportional to the beam current.
The total charge distribution along the beam direction shown in Fig. \ref{Raftari2015fig14}
demonstrates that, compared to the $160$~pA current,
the significant spatial variations happen closer to the sample-vacuum interface,
approximately within the depth of $40$~nm.
The system reaches the steady state around $0.1$~ms.
The densities of free electrons and holes reach the maxima of
$1.77\times10^{16}\ \text{cm}^{-3}$ and $4.5\times10^{17}\ \text{cm}^{-3}$, respectively.

This simulation confirms that a lower beam current results in significant differences
in the densities and spatial distribution between the free electrons and the free holes.
At lower beam current the free electrons tend to concentrate at the center of the impact zone
at some distance to the sample-vacuum interface, while the free holes are densely concentrated
at the interface (see Fig.~\ref{Raftari2015fig14}).

The lower-right image of Fig.~\ref{Raftari2015fig14} shows the spatial distribution of the
recombination term $U$, with the
highest rate of $8.25\times10^{24}\ \text{cm}^{-3}\text{s}^{-1}$ achieved in the steady state.
In the beginning of the process the recombination occurs at the center of the impact zone
and at the sample-vacuum interface, approximately at the same rate.
After a short time (few microseconds), the recombination is mostly active around the center
of the impact zone due to the low density of free electrons at the surface.
The highest recombination rates of $5.45\times10^{26}$ and $5.79\times10^{29}\ \text{cm}^{-3}\text{s}^{-1}$
in the steady state are obtained with the beam currents of 160 pA and 160 nA, respectively.
These results indicate that the recombination rate increases almost linearly
with the beam current.
Also, the variation of the recombination rate with the beam current clearly shows that
applying a constant recombination rate in computational models
does not reflect the actual recombination process in dielectric samples under SEM.

The SE yield can be calculated from the electron flux through the sample-vacuum interface
shown in Figures~\ref{Raftari2015fig8}, \ref{Raftari2015fig12}, and \ref{Raftari2015fig14}.
A clear discrepancy in the measured SE yield of insulators is found in the literature,
which can be attributed to the differences in the assumptions and conditions of the experiments.
The present simulations, where the only varying parameter is the applied beam current,
show that the SE yield increases with the beam current (provided all other conditions are fixed).
The SE yields obtained here for the particular $\text{SiO}_2$ sample under focused beam currents of $16$~pA,
$160$~pA, and $160$~nA are: $0.12$, $1.67$, and $2.74$, respectively,
which are, generally, within the range of experimental values reported in the literature.
In fact, we observe a weak (logarithmic) dependence of the SE yield on the beam current.
The experimental values of the SE yield for the $\text{SiO}_2$ (steam formed) sample
in the database by Joy \cite{Database} are: $0.25$, $1.02$, and $1.18$.
The measured SE yield for the $\text{SiO}_2$ (quartz) sample in \cite{barnardmeasurements}
is approximately $3$. Whereas, according to the experiment of Yong et al.
\cite{yong1998determination}, the SE yield of a ``wet and sputtered'' silicon dioxide
is found to be greater than~$3$.

\section*{Conclusions}
We have proposed a fully self-consistent drift-diffusion-reaction model 
augmented with a dynamic charge trapping model for the quantitative numerical 
investigation of the electron beam interaction with dielectric samples.
The pulsed and non-equilibrium nature of the charge injection mechanism and the back reaction
of the accumulated charge on the incoming primary electrons are incorporated
in the model through an explicit semi-empirical source formula. 
We have presented and compared two approaches to the charge injection problem.
The first one is a pulsed source model reflecting the actual discrete nature of 
the electron beam.  The second approach reduces the computational burden by applying 
a temporal average of the actual pulsed source function, which allows simulation at much 
longer time scales. Our results confirm the agreement between these two approaches in the 
initial stage and in the saturation regime. 

The proposed model features a Robin-type semi-permeable boundary
condition at the sample-vacuum interface reflecting the fact that the
electrons are allowed to go through the boundary, while holes are not.
We have devised a simple optimization procedure to deduce the surface
recombination velocity (SRV) of dielectrics in vacuum from the experimental SE yield data.

The results of our simulations clearly demonstrate the need for the dynamic trapping
equations in computational models of this kind. The trapping dynamics, namely,
the time evolution of the spatial distributions of the trapped charge densities,
has a major influence on (and helps to explain) the total charge distribution within
the sample and the apparent transients in the surface charge density (in the high-current regime).

Inclusion of the dynamic generation-recombination model is also deemed necessary,
since, as it turns out, the recombination rate depends on the beam current.
Other quantities depend either strongly or weakly on the beam
current as well, e.g., the local charge densities in
the steady state show a linear dependence, whereas, the dependence of the 
SE yield turns out to be logarithmic. 

Another conclusion of our study that requires a deeper mathematical analysis
is the apparent existence of a time-periodic steady state in the considered
system of equations for a pulsed source with PE impacts at regular intervals
observed in the neighborhood of the steady-state solution for an
averaged time-uniform source.

\section*{Acknowledgements}
The authors acknowledge the partial financial support of the FEI Company (Netherlands)
and many fruitful discussions with Dr.~S.~Sluyterman and Dr.~E.~Bosch (both with FEI).
Thanks are also due to Professor B.J. Thijsse (Delft University of Technology), who gave us valuable counsel
in the early stages of this work. Finally, we thank the anonymous reviewers 
for many thoughtful comments and suggestions.

%%%%%%%%%%%%%%%%%%%%%%%%%%%%%%%%%%%%%%%%%%%%%%%%%%%%%%%%%%%%%%%%%%%%%%%%%%%%%%%%%%%%%%%%%%%%%%%%%%%%%
\appendix\section{Weak form}\label{sec:app}
Consider the partial differential equations (\ref{equ:Po}), (\ref{equ:electron}),
and (\ref{equ:hole}) in the axisymmetric geometry (see Fig: \ref{Raftari2015fig2}):
\begin{align}
\label{equ:axisV}
\begin{split}
 &-\frac{1}{r}\frac{\partial}{\partial r}\left(\varepsilon r \frac{\partial V}{\partial r}\right)
 -\frac{\partial}{\partial z}\left(\varepsilon \frac{\partial V}{\partial z}\right)=Q(r,z),\\
 &Q(r,z)=\begin{cases}
              \frac{q}{\varepsilon_0}(p+p_t-n-n_t), & \text{in}\ \ \Omega_2\\
              0, & \text{in}\ \ \Omega_1
        \end{cases},
 \end{split}
\end{align}
\begin{align}
\label{equ:axisn}
\begin{split}
 &\frac{\partial n}{\partial t}-D_n\left(\frac{1}{r}\frac{\partial}{\partial r}
 \left(r\frac{\partial n}{\partial r}\right)+\frac{\partial^2 n}{\partial z^2}\right)\\
 &+\mu_n\left(\frac{\partial n}{\partial r}\frac{\partial V}{\partial r}+\frac{\partial n}{\partial z}\frac{\partial V}{\partial z} \right)
 +\mu_n n\left( \frac{1}{r}\frac{\partial}{\partial r}
 \left(r\frac{\partial V}{\partial r}\right)+\frac{\partial^2 V}{\partial z^2} \right)\\
 &=R+S_n-\frac{\partial n_t}{\partial t}\ \ \ \text{in}\ \ \Omega_2,
 \end{split}
\end{align}
\begin{align}
\label{equ:axisp}
\begin{split}
 &\frac{\partial p}{\partial t}-D_p\left(\frac{1}{r}\frac{\partial}{\partial r}
 \left(r\frac{\partial p}{\partial r}\right)+\frac{\partial^2 p}{\partial z^2}\right)\\
 &-\mu_p\left(\frac{\partial p}{\partial r}\frac{\partial V}{\partial r}+\frac{\partial p}{\partial z}\frac{\partial V}{\partial z} \right)
 -\mu_p p\left( \frac{1}{r}\frac{\partial}{\partial r}
 \left(r\frac{\partial V}{\partial r}\right)+\frac{\partial^2 V}{\partial z^2} \right)\\
 &=R+S_p-\frac{\partial p_t}{\partial t}\ \ \ \ \text{in}\ \ \Omega_2.
 \end{split}
\end{align}

To derive the weak formulation we integrate over the cross-sectional area ($rdrdz$) arriving
at the following form of the equation (\ref{equ:axisV}):
\begin{align}
 \int_{\Omega_1\cup \Omega_2}w\left( -\frac{1}{r}\frac{\partial}{\partial r}\left(\varepsilon r \frac{\partial V}{\partial r}\right)
 -\frac{\partial}{\partial z}\left(\varepsilon \frac{\partial V}{\partial z}\right) -Q\right)rdrdz=0,
\end{align}
where $w$ is the weight function. Integrating the highest-order terms by parts we obtain:
\begin{align}
\begin{split}
 &-\int_{\Omega_1\cup \Omega_2}w \frac{\partial}{\partial r}\left(\varepsilon r \frac{\partial V}{\partial r} \right)drdz= \\
 &\int_{\Omega_1\cup \Omega_2}\varepsilon \frac{\partial w}{\partial r}\frac{\partial V}{\partial r} rdrdz
 -\int_{\partial\Omega_1\cup \partial\Omega_2}\varepsilon w\frac{\partial V}{\partial r} r \hat{\nu}_rds,
 \end{split}
\end{align}
\begin{align}
\begin{split}
 &-\int_{\Omega_1\cup \Omega_2}w \frac{\partial}{\partial z}\left(\varepsilon \frac{\partial V}{\partial z} \right)rdrdz= \\
 &\int_{\Omega_1\cup \Omega_2}\varepsilon \frac{\partial w}{\partial z}\frac{\partial V}{\partial z} rdrdz
 -\int_{\partial\Omega_1\cup \partial\Omega_2}\varepsilon w\frac{\partial V}{\partial z}r \hat{\nu}_zds,
 \end{split}
\end{align}
where $\hat{\nu}=\langle\hat{\nu}_r,\hat{\nu}_z\rangle$ is the outward unit normal vector to the boundary.
Therefore, the weak form of the equation (\ref{equ:axisV}) can be written as:
\begin{align}
\begin{split}
&\int_{\Omega_1\cup \Omega_2}\varepsilon \frac{\partial w}{\partial r}\frac{\partial V}{\partial r} rdrdz
 =\int_{\Omega_1\cup \Omega_2}\varepsilon \frac{\partial w}{\partial z}\frac{\partial V}{\partial z} rdrdz\\
 &-\int_{\Omega_1\cup \Omega_2}wQrdrdz=0,
 \end{split}
\end{align}
or
\begin{align}
 \int_{\Omega_1\cup \Omega_2}\varepsilon (\nabla w \cdot \nabla V ) rdrdz
 -\int_{\Omega_1\cup \Omega_2}wQrdrdz=0,
\end{align}
where $w\mid_{\partial\Omega_1\cup\partial\Omega_2}=0$ and
$\nabla=\langle\partial/\partial r, \partial/\partial z\rangle$.

The weak form of the equation (\ref{equ:axisn}) is:
\begin{align}
\begin{split}
  &\int_{\Omega_2}  w\Bigg(\frac{\partial n}{\partial t}-D_n\left(\frac{1}{r}\frac{\partial}{\partial r}
 \left(r\frac{\partial n}{\partial r}\right)+\frac{\partial^2 n}{\partial z^2}\right)\\
 &+\mu_n\left(\frac{\partial n}{\partial r}\frac{\partial V}{\partial r}+\frac{\partial n}{\partial z}\frac{\partial V}{\partial z} \right)\\
  &+\mu_n n\left( \frac{1}{r}\frac{\partial}{\partial r}
 \left(r\frac{\partial V}{\partial r}\right)+\frac{\partial^2 V}{\partial z^2} \right)\\
 &-R-S_n+\frac{\partial n_t}{\partial t}\Bigg)rdrdz=0,
 \end{split}
\end{align}
Integrating the highest-order terms by parts we get:
\begin{align}
\begin{split}
 &\int_{\Omega_2}w \left(\frac{1}{r}\frac{\partial}{\partial r}
 \left(r\frac{\partial n}{\partial r}\right)+\frac{\partial^2 n}{\partial z^2}\right)rdrdz=\\
 &-\int_{\Omega_2} \left(\frac{\partial w}{\partial r}\frac{\partial n}{\partial r}+
 \frac{\partial w}{\partial z}\frac{\partial n}{\partial z}\right) rdrdz\\
 &+\int_{\Sigma \cup \partial\Omega_2} rw \left(\frac{\partial n}{\partial r}\hat{\nu_r}
 +\frac{\partial n}{\partial z}\hat{\nu_z}\right)ds,
\end{split}
\end{align}
\begin{align}
\begin{split}
 &\int_{\Omega_2}w n\left(\frac{1}{r}\frac{\partial}{\partial r}
 \left(r\frac{\partial V}{\partial r}\right)+\frac{\partial^2 V}{\partial z^2}\right)rdrdz=\\
 &-\int_{\Omega_2} \left(\frac{\partial (wn)}{\partial r}\frac{\partial V}{\partial r}+
 \frac{\partial (wn)}{\partial z}\frac{\partial V}{\partial z}\right) rdrdz\\
 &+\int_{\Sigma \cup \partial\Omega_2} w n\left(\frac{\partial V}{\partial r}\hat{\nu_r}
 +\frac{\partial V}{\partial z}\hat{\nu_z}\right)rds=\\
 &-\int_{\Omega_2} n\left(\frac{\partial (w)}{\partial r}\frac{\partial V}{\partial r}+
 \frac{\partial (w)}{\partial z}\frac{\partial V}{\partial z}\right) rdrdz\\
 &-\int_{\Omega_2} w\left(\frac{\partial (n)}{\partial r}\frac{\partial V}{\partial r}+
 \frac{\partial (n)}{\partial z}\frac{\partial V}{\partial z}\right) rdrdz\\
 &+\int_{\Sigma \cup \partial\Omega_2} w n\left(\frac{\partial V}{\partial r}\hat{\nu_r}
 +\frac{\partial V}{\partial z}\hat{\nu_z}\right)rds.
\end{split}
\end{align}
Therefore, the weak form of the equation (\ref{equ:axisn}) can be written as:
\begin{align}
\begin{split}
&\int_{\Omega_2}  w\left(\frac{\partial n}{\partial t}-R-S_n+\frac{\partial n_t}{\partial t}\right)rdrdz\\
&-\int_{\Omega_2} \nabla w \cdot (-D_n\nabla n+\mu_nn\nabla V) rdrdz\\
&+\int_{\Sigma \cup \partial\Omega_2}w(-D_n\nabla n+\mu_nn\nabla V)\cdot \hat{\nu}\ rds=0.
\end{split}
\end{align}
Substitution of the boundary conditions gives:
\begin{align}
\begin{split}
&\int_{\Omega_2}  w\left(\frac{\partial n}{\partial t}-R-S_n+\frac{\partial n_t}{\partial t}\right)rdrdz\\
&-\int_{\Omega_2} \nabla w \cdot (-D_n\nabla n+\mu_nn\nabla V) rdrdz\\
&+\int_{\Sigma}wv_n(n-n_i)\ rds=0,
\end{split}
\end{align}
where $n\mid_{\partial\Omega_2}=n_i$ and $w\mid_{\partial\Omega_2}=0$.

Along similar lines the weak form of the equation (\ref{equ:axisp}) cab be derived as:
\begin{align}
\begin{split}
&\int_{\Omega_2}  w\left(\frac{\partial p}{\partial t}-R-S_p+\frac{\partial p_t}{\partial t}\right)rdrdz\\
&-\int_{\Omega_2} \nabla w \cdot (-D_p\nabla p-\mu_pp\nabla V) rdrdz=0
\end{split}
\end{align}
where $p\mid_{\partial\Omega_2}=n_i$ and $w\mid_{\partial\Omega_2}=0$.

% \bibliographystyle{plain}
% \bibliography{mybib}

\end{document}